\documentclass[prl,aps,twocolumn,showpacs]{revtex4}

\usepackage{amsfonts}
\usepackage{amssymb}
\usepackage{graphicx}
\usepackage{bbold}
\usepackage{textcomp}

\usepackage{bm}

\begin{document}
\title{Heralded photon amplification for quantum communication}\author{C.~I.~Osorio, N.~Bruno, N.~Sangouard, H.~Zbinden, N.~Gisin, R.~T.~Thew }\affiliation{Group of Applied Physics, University of Geneva,
1211 Geneva 4, Switzerland}

\begin{abstract}
Heralded noiseless amplification based on single-photon sources and linear optics is  ideally suited for long-distance quantum communication tasks based on discrete variables. We experimentally demonstrate such an amplifier, operating at telecommunication wavelengths. Coherent amplification is performed with a gain of $G=1.98\pm0.2$, for a state with a maximum expected gain $G=2$. We also demonstrate that there is no need for a stable phase reference between the initial signal state and the local auxiliary photons used by the amplifier. These results highlight the potential of heralded quantum amplifiers for long-distance quantum communication, and bring device-independent quantum key distribution one step closer.
\end{abstract}



\maketitle


Transmission loss is a fundamental limitation in quantum communication. A photon $|1\rangle$ propagating in a channel with a transmission efficiency $\eta_t,$ ends up in a state given by $\eta_t|1\rangle\langle 1| + (1-\eta_t) |0\rangle\langle 0|$. Amplifying the single-photon  component $|1\rangle$ is not possible in a deterministic, noiseless and coherent operation. In fact, if such an operation would be possible it could be used for signaling \cite{Gisin98}. This restriction makes long-distance quantum communication very challenging, but it also offers a unique opportunity for key distribution with an unequaled security \cite{Gisin02}.

Even when the amplification of a quantum state cannot be deterministically performed with unit fidelity,  approximate quantum amplification is possible. Probabilistic noiseless amplifiers have recently attracted a lot of attention. Of particular interest are those where the success of the amplification process is heralded. This includes techniques based on single-photon addition \cite{Fiurasek09, Zavatta11}, or thermal noise addition followed by heralded photon subtraction \cite{Marek10, Usuga10}.

Reference~\cite{Ralph09} presents an interesting alternative for realizing heralded quantum amplification. Inspired by the concept of quantum scissors \cite{Pegg81}, the authors propose a scheme requiring only single-photon sources and linear optics. This is an attractive proposal from a practical point of view, and it has already triggered  a couple of proof-of-principle experiments~\cite{Xiang10, Ferreyrol10}. These experiments focus on the applications of the amplifiers in continuous variable based quantum information science, e.g. distilling  continuous variable entanglement \cite{Xiang10}, or improving continuous-variable quantum key distribution \cite{Ferreyrol10, Fossier09}.

\begin{figure}[thb!]
\centering
\includegraphics[scale=0.60]{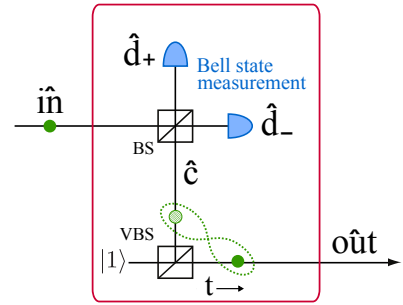}
\caption{A heralded single photon amplifier. The auxiliary photon $|1\rangle$ generates an entangled state between the modes $\rm{\hat{c}}$ and $\rm{\hat{out}}$ after passing through  a variable beamsplitter (VBS) with transmission $t$. The input mode $\rm{\hat{in}}$ is then combined with  $\rm{\hat{c}}$ in a $50/50$ beamsplitter (BS). When $t$= 50\%,  a measurement in $\rm{\hat{d}_{+}}$ or $\rm{\hat{d}_{-}}$ produces a teleportation transformation between the $\rm{\hat{in}}$ and $\rm{\hat{out}}$ modes \cite{Boschi98,Lombardi02}. When $t>50\%$ such a detection heralds the desired noiseless amplification.}
\label{fig:amplith}
\end{figure}

\begin{figure*}[htb!]
\centering
\includegraphics[height=50mm]{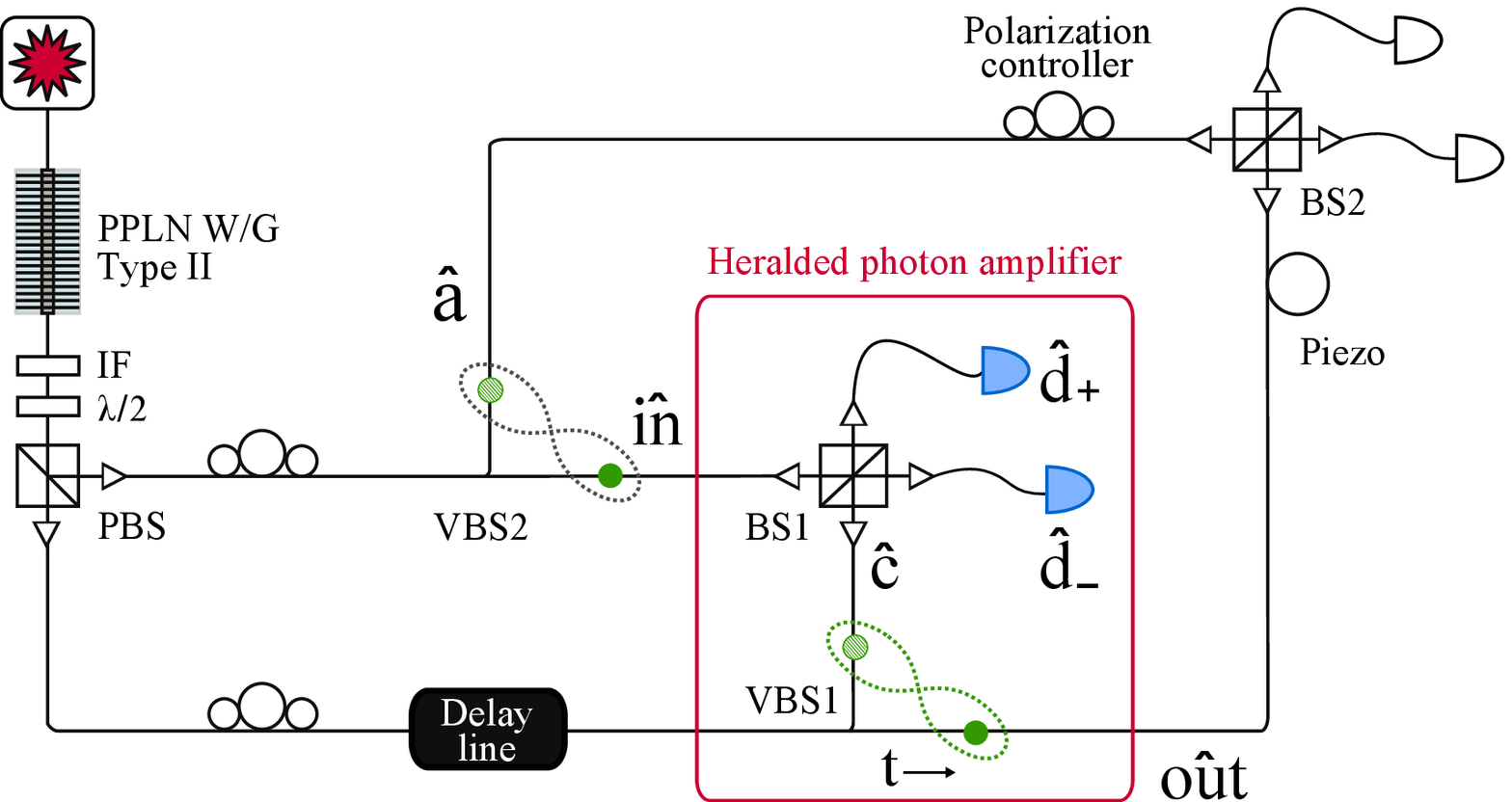}\hspace{5mm}
\includegraphics[height=45mm]{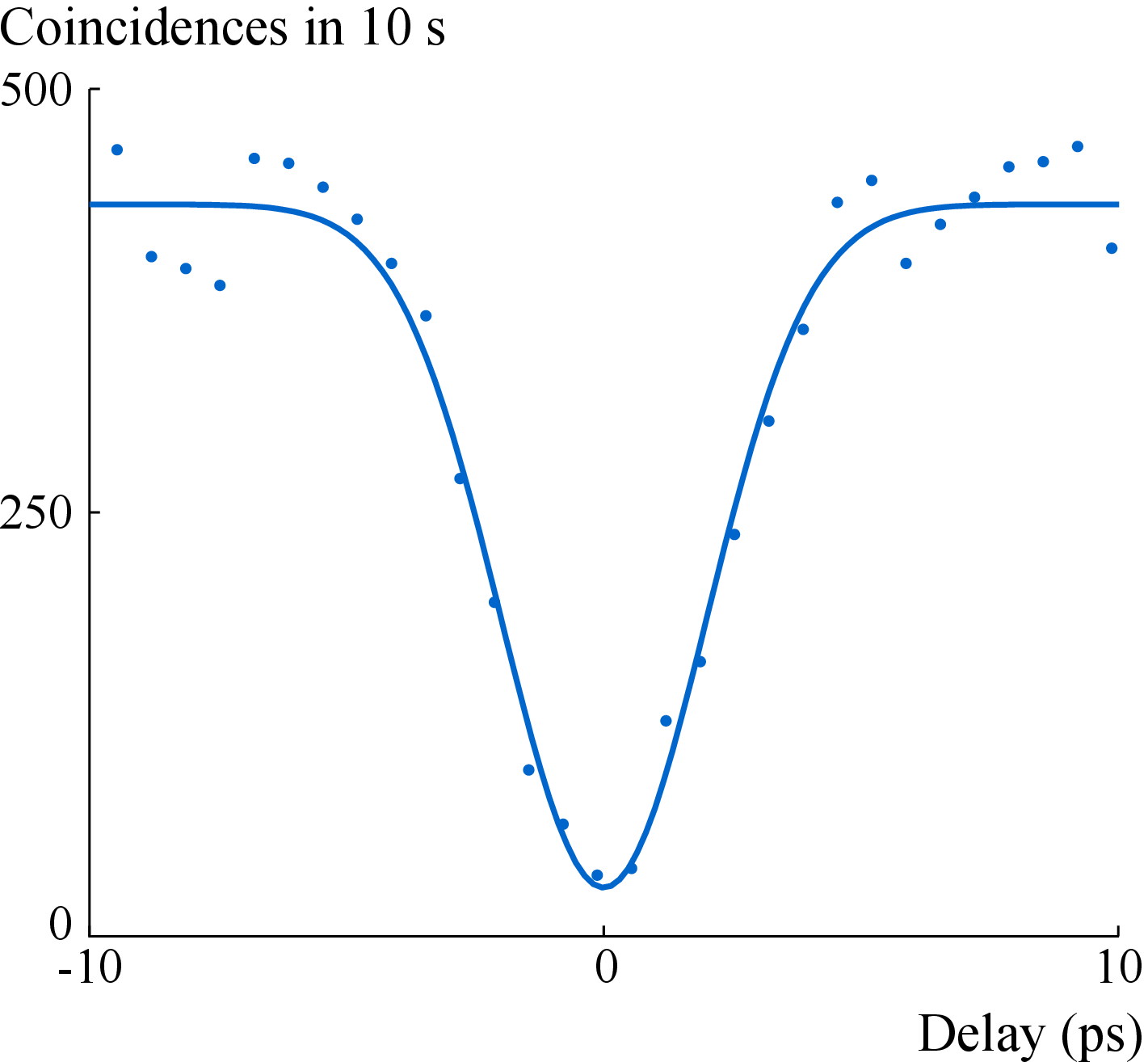}
\caption{(Left) A schematic of the set-up used to demonstrate heralded single photon amplification.  Two photons, produced via SPDC, are fed into the heralded photon amplifier, which is embedded in an interferometer formed by two variable fiber beamsplitters (VBS), and two bulk  $50/50$ beamsplitters (BS). (Right) The high visibility of the HOM interference, measured after BS1 \& BS2, proves the spectral and temporal indistinguishability of the photons.  The figure shows the measured points, as well as the fit used to estimate the visibility ($93.4\pm 5.9\%$) after BS1.}
\label{fig:schematicanddip}
\end{figure*}

Unlike recent realizations \cite{Xiang10, Ferreyrol10}, we focus on the potential of the amplifier in Ref.~\cite{Ralph09} for tasks based on discrete variables. Our experiment uses photons at telecom wavelengths, so it is ideally suited for long-distance quantum communication. Furthermore, since it is based on polarization independent elements, our device could also be used as a qubit amplifier, opening up the way for experimental  device-independent quantum key distribution (DI-QKD)  \cite{Gisin10}.

In our system we have direct access to the behavior of the gain, and we also demonstrate the coherence preserving nature of the process.  Additionally, we show that no specific phase stability is needed between the input photon and the auxiliary photon, as required in a distributed quantum network \cite{Minar11}.

This letter is divided into three sections: The first  describes the theoretical aspects of our single photon amplifier; the second  explains the experimental set-up, the measurements performed and the results obtained, and the final section discusses the additional requirements to implement the amplifier in a real quantum network.


\section{Theory}
Figure \ref{fig:amplith} depicts a heralded single photon amplifier. We consider an input state that is a statistical mixture between the vacuum and a single photon

\begin{equation}
\label{initialstate}
\alpha^2 \-\ |0\rangle\langle 0| +\beta^2 \-\  \rm{\hat{in}}^\dag |0\rangle \langle 0| \rm{\hat{in}}.
\end{equation}

\noindent An auxiliary photon, sent through a variable beamsplitter (VBS) with transmission $t$, produces the entangled state between the modes $\rm{\hat{c}}$ and $\rm{\hat{out}}$ given by

\begin{equation}
\sqrt{1-t} \-\ \rm{\hat{c}}^\dag|0\rangle + \sqrt{t} \-\ \rm{\hat{out}}^\dag|0\rangle.
\end{equation}

\noindent Then, a 50/50 beamsplitter combines   $\rm{\hat{c}}$ and $\rm{\hat{in}}$ producing the modes $\rm{\hat{d}_{+}}=\frac{1}{\sqrt{2}}(\rm{\hat{c}}+\rm{\hat{in}})$ and $\rm{\hat{d}_{-}}=\frac{1}{\sqrt{2}}(\rm{\hat{c}}-\rm{\hat{in}})$.  If the auxiliary and the input photon are indistinguishable, the detection of a single photon in one of these modes, say in the mode $\rm{\hat{d}_{+}}$, acts as a Bell state measurement and projects the output state $\rm{\hat{out}}$ into

\begin{equation}
\label{out}
\frac{1}{N}\left[(1-t)\alpha^2 \-\ |0\rangle\langle 0|+ t \beta^2 \-\  \rm{\hat{out}}^\dag |0\rangle \langle 0| \rm{\hat{out}}\right].
\end{equation}

\noindent Where $N=(1-t)\alpha^2+t\beta^2$ is a normalization factor.  Since the detection of one photon in $\rm{\hat{d}_{-}}$ also leads to Eq.~(\ref{out}), the probability to successfully produce that state is given by twice its norm.

The process of transforming the input into the output state changes the weight of the single-photon component. The gain $G$ of the process quantifies the change, it is defined as the ratio of the single-photon component weights before and after the process, and it is given by $G = \frac{t}{t+\alpha^2 (1 - 2t)}$. Importantly, if there is a well defined phase between the single-photon and the vacuum component on  Eq.~(\ref{initialstate}), it is preserved by the process. Therefore, in the case where $t=\frac{1}{2}$, the input state is mapped to the output, the gain $G=1$, and the protocol reduces to the teleportation  of a vacuum-single-photon qubit \cite{Boschi98,Lombardi02}.  When $t>\frac{1}{2}$ the gain $G>1$ and the desired noiseless amplification is realized.

The amplification, as explained before, is always heralded by a signal coming from either the detection of a single photon in  $\rm{\hat{d}_{+}}$ or  $\rm{\hat{d}_{-}}$.  However, in the case of non photon-number resolving detectors it can be shown that the gain reduces to $G=\frac{t}{1-\alpha^2 t}$. For instance, for $\alpha^2=1/2$ the amplification only takes place for $t>\frac{2}{3}$. In practice, the gain depends on  myriad parameters that affect the output state given by Eq.~(\ref{out}). To describe our experimental amplifier, we developed a model for the gain taking into account the coupling, transmission losses and detector efficiency, which can be quantified independently.


\section{Experiment}
Figure \ref{fig:schematicanddip} shows a schematic of the experimental set-up.  A $25$\,mm long type-II PPLN waveguide  (University of Paderborn) generates both the input and the auxiliary photon through spontaneous parametric down conversion.  The interaction of the waveguide with a continuous-wave diode laser at $780$\,nm (Toptica DL100, $3.3$ mW) results in the creation of degenerated photon pairs at $1560$\,nm.  After the waveguide,  a
silicon filter blocks the laser, and an interference filter (IF) with $1.3$\,nm bandwidth guarantees the spectral indistinguishability of the photons. The pairs are separated by a polarization beamsplitter (PBS) and then coupled into a single mode optical fibers, which ensures their spatial indistinguishability.   A delay line in one of the fibers controls the path length difference.

\begin{figure}[tbh!]
\includegraphics[scale = 0.40]{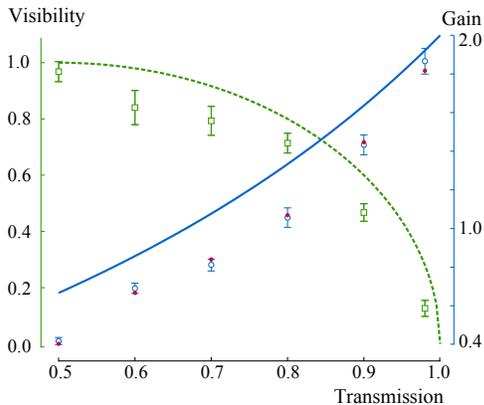}
\caption{Measured values of the fidelity $V$ and gain $G$ as a function of the transmission $t$ (squares and circles respectively).  The dashed green line is the maximum possible visibility $V_{max}$. The solid blue line is the maximum gain with non photon-number resolving detectors, without losses and with $\alpha^2=1/2$, the full circles are the values estimated by our model.}
\label{fig:results}
\end{figure}

The heralded photon amplifier is embedded in an interferometer formed by two variable fiber beamsplitters (VBS), and two bulk  $50/50$ beamsplitters (BS).  While VBS1 is part of the amplifier scheme, VBS2 is used to prepare the input state. An input photon passing through VBS2 produces an entangled state between the modes $\rm{\hat{a}}$ and $\rm{\hat{in}}$. When considered independently, the state of each of those modes is given by Eq.~(\ref{initialstate}) (with  $\alpha$ and $\beta$ set by the splitting ratio of VBS2). We use $\rm{\hat{in}}$ as input mode and, as we will explain later, we use $\rm{\hat{a}}$ to test the coherence of the process by combining it with the mode $\rm{\hat{out}}$ at BS2.

The outputs of the two bulk beamsplitters BS1 and BS2 are coupled into single mode fibers so that they can be connected to the detectors. The hybrid configuration of bulk and fiber optics, although not necessary for the amplification, has been adopted to facilitate the measurements needed to quantify the gain.

The two-photon coincidence events were measured via two InGaAs/InP avalanche photodiodes. We measure single-photon events using a free running detector (IDQ - ID210 with $10\,\%$ efficiency, $3$\,KHz noise and $20$\,$\mu$s dead time). This detection triggers a second detector in gated mode that measures the coincidences (IDQ - ID200 with $15\,\%$ efficiency, $2.5$\,ns detection gate, and a dark count probability of $3\times10^{-5}$ per detection gate).

The amplification process requires that the input and auxiliary photon are indistinguishable in every degree of freedom.  To quantify the indistinguishability, we  first measure the Hong-Ou-Mandel (HOM) interference on each beamsplitter (BS1 \& BS2). The interferometer is made such that the path difference between the inputs of both beamsplitters is equal (within the coherence length of the photons). The coincidences after BS1 \&  BS2 were measured as a function of the delay between the input photons.   The characteristic dips had visibilities of $93.4\pm 5.9\%$  and $92.1\pm5.7\%$ for BS1 and BS2 respectively. The right hand side of Fig. \ref{fig:schematicanddip} shows the results for BS1.

\begin{figure}[thb!]
\includegraphics[scale = 0.40]{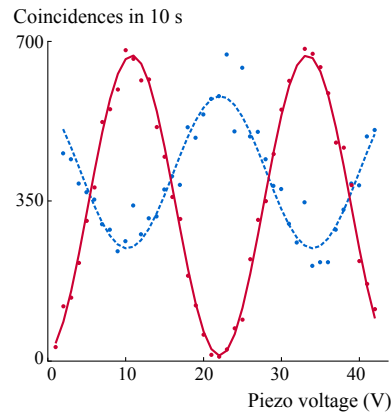}
\caption{Coincidences between the heralding detector at  ${\rm \hat{d}_{-}}$ and one output of BS2 as the phase is varied, with $t=50\%$.  For the amplifier (solid red curve), the high visibility of the fringes ($96.8\pm3.5\%$) shows that coherence is preserved in the protocol. The dashed blue curve is obtained with an input state given by  Eq.~ (\ref{phasestate}). The decrease on the visibility ($40.9\pm2.4\%$) indicates rapid fluctuations in the phase between the input modes of the interferometer.}
\label{fig:fringes}
\end{figure}

Even with indistinguishable photons,  double pair emission and detector imperfections decrease the visibility of a HOM dip.  Based on the model presented in Ref. \cite{Sekatski11}, we calculated a maximum visibility for our source of $92.7\%$ \cite{source}. The measured visibilities correspond then to highly indistinguishable photons.

To quantify the gain of the heralded amplifier we compare the probability $P_{in}$ of detecting a photon at $\rm{\hat{in}}$  with the probability $P_{out}$ of detecting a heralded photon in the mode $\rm{\hat{out}}$. These probabilities are obtained in two independent coincidence measurements. The first is made by blocking the paths $\rm{\hat{a}}$ and $\rm{\hat{c}}$ of the interferometer, and using the free running detector after BS2 and the second detector after BS1. The coincidences to singles ratio ($C_1/S_1$) provides our initial reference probability $P_{in}$ (approx. $15\times10^{-4}$ per detection gate) \cite{pin}.  In the second measurement only the path $\rm{\hat{a}}$ of the interferometer is blocked, and the positions of the detectors are reversed.  In this case, the coincidences to singles ratio ($C_2/S_2$) gives the amplified output probability $P_{out}$. We can then define the gain, in terms of the measured quantities, as

\begin{eqnarray}
G = P_{out}/P_{in}.
\label{Gexp}
\end{eqnarray}

\noindent Figure \ref{fig:results} shows the gain measured for different values of the transmission $t$ of VBS1, where the splitting ratio of VBS2 was fixed at $50/50$. Then, the input state is then given by Eq.~(\ref{initialstate}) with $\alpha^2=\beta^2=1/2$, and therefore the maximum possible gain is $G=2$ for $t=1$. In our system the amplification $G > 1$ occurs for $t >75\%$, and the gain reaches a value of $G=1.98\pm0.2$ at $t=98\%$.  The measured values of the gain are in good agreement with those predicted taking into account the coupling, transmission losses and detector efficiency.

In order to fully demostrate amplification, for each gain measurement it is necessary to verify the coherence preserving nature of the process. To do so we can take advantage of the fact that for $t=50\%$ our experiment is analogous to an entanglement swapping experiment \cite{Sciarrino02}. Following Fig. \ref{fig:schematicanddip}, we generate two maximally entangled states after VBS1 and VBS2. Then, a Bell state measurement at BS1  swaps the entanglement onto the two modes $\rm{\hat{a}}$  and $\rm{\hat{out}}$.

With $\rm{\hat{a}}$  and $\rm{\hat{c}}$ unblocked,  we measured the coincidences between the heralding detector at  ${\rm \hat{d}_{-}}$ and one output of BS2, as the phase between the inputs of BS2 is varied. To control this phase, part of the interferometer's fiber is wrapped around a piezoelectric tube.   Figure \ref{fig:fringes} shows the interference pattern obtained in the coincidences for $t=50\%$. The interference clearly indicates that the coherence is preserved during the process.

At higher values of $t$, due to an increasing imbalance in the interferometer,  the expected value of the interference visibility $V_{max}$  decreases following the expression $V_{max}= 2\sqrt{t(1-t)}$ \cite{visibility}.  Figure \ref{fig:results} shows the measured values of the visibility as well as $V_{max}$ for different values of $t$.

If we are to use the amplifier in a distributed quantum network  it is important for the protocol to be independent of the phase between the different sources.   Therefore, in addition to the measurements of  gain and coherence, we studied the phase relation between the input modes of the interferometer.   During our experiment   this phase fluctuates rapidly (the fibers between the source and the interferometer are not stabilized). The effect of this fluctuation is clear if we turn a $\lambda / 2$ wave plate (before the PBS in Fig. \ref{fig:schematicanddip}) in order  to have an input state of the form

\begin{eqnarray}
 |\psi\rangle = \frac{1}{\sqrt{2}}e^{i \phi}|1,1\rangle + \frac{1}{2}|2,0\rangle + \frac{1}{2}e^{2i \phi}|0,2\rangle,
\label{phasestate}
\end{eqnarray}

\noindent where $|n,m\rangle$ are the the input modes of VBS1 and VBS2, and  $\phi$ is a phase between these two modes.  Figure \ref{fig:fringes} compares the interference fringes obtained for the amplifier, with those obtained with the input state given by  Eq.~(\ref{phasestate}).  In the second case  the $|1,1\rangle$ component interferes as before, but the rapid fluctuations of $\phi$ erase the interference of the last two terms.
Figure \ref{fig:fringes} shows how the visibility decreases by a factor of $2$, as expected. Therefore, in our scheme the phase between the input state and the auxiliary photon is not stable, which does not affect the amplification process.

\section{Discussion}
In contrast to our amplifier, where one source generates both photons, in a real quantum network the input state and the auxiliary photon would be independently generated. That is also the case for the implementation of a qubit amplifier \cite{Gisin10}.  With independently generated photons the interference requires purity in addition to indistinguishability.  While the indistinguishability can be achieved as shown here,  to obtain photons in pure states is challenging. Strong filters may remove the correlations but they also reduce the amount of available photons. A more convenient solution is to obtain pure photons directly from the source. There has been a large effort to design and implement pure photon sources as reported, for example, in Ref. \cite{Mosley08,Levine10,Svozilik11,Pomarico11}. However, it is still a challenge to build a single-photon source satisfying all the requirements on frequency, bandwidth, and maximum transmission losses necessary for an amplifier in a real quantum network.

\section{Conclusions}
We have reported in this letter the successful heralded amplification of a single photon in the telecommunication regime. We have shown a maximum gain of $\sim$ $2$ and demonstrated the coherence of this process. We have also shown that the amplification scheme does not require a stable phase between the input state and the auxiliary photon. All of these results highlight the potential of heralded quantum amplifiers in long-distance quantum communication based on quantum repeaters \cite{Minar11}. The experiment reported here also brings, for the first time, device-independent quantum key distribution into the realm of experimental physics \cite{Gisin10}.

We would like to thank W. Solher and the Integrated Optics Group in the University of Paderborn for the development and fabrication of the custom PPLN waveguide. We also acknowledge useful discussions with C. Branciard, and financial support from the EU project Q-essence and the Swiss project NCCR-QSIT.


\begin{thebibliography}{99}
\bibitem{source}  The source emits $\approx 1\times10^{-4}$ pairs/ps, and  the coincidence detections have  a jitter of $400$\,ps.
\bibitem{pin} Since the coincidences are measured after BS1 and BS2, there is a factor of $2$ for both $P_{in}$ and $P_{out}$ that cancels out in Eq.~(\ref{Gexp}).
\bibitem{visibility} In an independent measurement, we canceled the effect of the imbalance by changing the splitting ratio of VBS2. At $t=90\%$, changing the splitting ratio of VBS2 from $50/50$ to $90/10$, increases the visibility to $96.3\pm3.8\%$ (from $46.7\pm3.1\%$).
\end{thebibliography}
\end{document}